\documentclass[acmtog]{acmart}

\usepackage{algorithm}
\usepackage{algorithmic}
\usepackage{subcaption}
\usepackage{multirow}
\usepackage{colortbl}

\makeatletter
\renewcommand*{\ALG@name}{Procedure}
\makeatother
\makeatletter
\newcommand{\ALC@uniqueautorefname}{Line}
\makeatother

\definecolor{FirstPlace}{RGB}{255,200,200}  
\definecolor{SecondPlace}{RGB}{255,225,180} 
\definecolor{ThirdPlace}{RGB}{255,255,180}  

\graphicspath{ {./image/} }

\AtBeginDocument{%
  }

\setcopyright{acmlicensed}
\copyrightyear{2018}
\acmYear{2018}
\acmDOI{XXXXXXX.XXXXXXX}

\acmJournal{TOG}
\acmSubmissionID{2107}

\begin{document}

\title{Efficient Differentiable Hardware Rasterization for 3D Gaussian Splatting}

\author{Yitian Yuan}
\affiliation{%
  \institution{Shanghai Jiao Tong University}
  \city{Shanghai}
  \country{China}
}
\email{y13916619121@126.com}

\author{Qianyue He}
\affiliation{%
  \institution{Tsinghua University}
  \country{China}
}
\email{he-qy22@mails.tsinghua.edu.cn}


\begin{abstract}
  Recent works demonstrate the advantages of hardware rasterization for 3D Gaussian Splatting (3DGS) in forward-pass rendering through fast GPU-optimized graphics and fixed memory footprint. However, extending these benefits to backward-pass gradient computation remains challenging due to graphics pipeline constraints. We present a differentiable hardware rasterizer for 3DGS that overcomes the memory and performance limitations of tile-based software rasterization. Our solution employs programmable blending for per-pixel gradient computation combined with a hybrid gradient reduction strategy (quad-level + subgroup) in fragment shaders, achieving over 10$\times$ faster backward rasterization versus naive atomic operations and 3$\times$ speedup over the canonical tile-based rasterizer. 
  Systematic evaluation reveals 16-bit render targets (float16 and unorm16) as the optimal accuracy-efficiency trade-off, achieving higher gradient accuracy among mixed-precision rendering formats with execution speeds second only to unorm8, while float32 texture incurs severe forward pass performance degradation due to suboptimal hardware optimizations.
  Our method with float16 formats demonstrates 3.07$\times$ acceleration in full pipeline execution (forward + backward passes) on RTX4080 GPUs with the MipNeRF 360 dataset, outperforming the baseline tile-based renderer while preserving hardware rasterization's memory efficiency advantages --- incurring merely 2.67\% of the memory overhead required for splat sorting operations. This work presents a unified differentiable hardware rasterization method that simultaneously optimizes runtime and memory usage for 3DGS, making it particularly suitable for resource-constrained devices with limited memory capacity.

\end{abstract}


\received{20 July 2025}
\received[revised]{12 August 2025}
\received[accepted]{5 September 2025}

\maketitle

\section{Introduction}

3D Gaussian Splatting (3DGS) \cite{kerbl20233d} has established itself as an effective differentiable scene representation, demonstrating remarkable capabilities in multiview 3D reconstruction, novel view synthesis, and real-time rendering.

The differentiable renderer constitutes the core component of 3DGS-based scene reconstruction. It computes gradients of Gaussian splat parameters with respect to photometric loss functions, enabling iterative parameter optimization. Current implementations primarily rely on tile-based software rasterization, which operates through three key stages: (1) identifying tiles intersected by each Gaussian splat, (2) sorting splat-tile pairs which requires dynamic memory allocation due to the unpredictable number of intersections, and (3) executing two distinct processing phases: a forward pass for color computation via alpha blending and a backward pass for gradient propagation.

For the forward pass of the differentiable 3DGS rendering process, recent work has demonstrated substantial performance improvements over tile-based methods through hardware rasterization pipelines \cite{bulo2025hardware, kevin2023splat, fast_gauss}. The hardware-accelerated approach executes through three stages: (1) depth-sorting all splats, (2) rendering each splat as a quad bounded by its projected 2D covariance\cite{zwicker2002ewa}, and (3) computing per-pixel transparency $\alpha$ through fragment shaders based on the 2D Gaussian distribution, with hardware blending handling color composition. Crucially, this pipeline eliminates the dynamic memory management requirements of tile-based methods, replacing variable-size splat-tile buffers with fixed-capacity buffers scaled to the total splat count, thereby ensuring memory efficiency.

Although hardware rasterization has demonstrated significant performance and memory efficiency advantages for the forward pass in 3DGS rendering, extending these benefits to the backward pass presents a fundamental challenge: computing per-pixel splat gradients within functionality-constrained fragment shaders of the GPU graphics pipeline.
Meanwhile, efficiently accumulating these gradients per-splat remains an additional hurdle. Both challenges are unresolved in current implementations.

This paper thus introduces a novel differentiable 3DGS hardware rasterizer that overcomes these challenges. Our first contribution is a mathematically rigorous backward gradient propagation method that fully conforms to hardware rasterization limitations, which is achieved through GPU programmable blending technique to maintain pixel-local data required for gradient computations.

Our second contribution is an efficient splat-wise gradient reduction strategy in fragment shaders that combines quad-level ($2\times2$ pixel) and subgroup operations. This approach achieves 10$\times$ faster backward rasterization compared to baseline \texttt{atomicAdd} implementations and demonstrates a 3$\times$ speedup over the original tile-based backward rasterization \cite{kerbl20233d}.

To mitigate float32 render targets' performance overhead in forward pass, we evaluate reduced-precision alternatives (float16, unorm16, unorm8) and further develop mixed-precision rendering techniques. Our systematic analysis identifies float16 and unorm16 as the optimal formats, achieving relatively minor gradient computation errors (RMSE and MRE) while delivering near-peak throughput, surpassed only by our unorm8 configuration in raw speed.

On an RTX4080 GPU with the MipNeRF dataset\cite{barron2022mip}, our float16-based implementation achieves 3.07$\times$ faster end-to-end performance compared to the tile-based rasterizer proposed by \citet{kerbl20233d}. By adopting hardware rasterization with fixed-capacity buffers, we reduce memory consumption for splat sorting to just 2.67\% of tile-based approaches, simultaneously improving both computational speed and memory efficiency.

\section{Background and Related Work}

\subsection{3D Gaussian Splatting}

\subsubsection{Scene Parameterization}

\citet{kerbl20233d} proposed 3D Gaussian Splatting (3DGS) as a differentiable scene representation that models 3D geometry using anisotropic 3D Gaussians. Each Gaussian primitive is parameterized by: (1) a center position $\mu_\text{3D} \in \mathbb{R}^3$, (2) opacity $o \in [0,1]$, (3) view-dependent color encoded through spherical harmonics (SHs), and (4) a covariance matrix $\mathbf{\Sigma_\text{3D}}$ constructed from the learnable rotation parameters $\mathbf{R}$ and scaling parameters $\mathbf{S}$ as $\mathbf{\Sigma_\text{3D}} = \mathbf{RS}\mathbf{S}^\top\mathbf{R}^\top$. The influence of a Gaussian (i.e. density) at point $\mathbf{x}$ follows: 
\begin{equation}\label{eqn:3dgs-basic}
 \alpha_\text{3D}(\mathbf{x}|o,\mu_{\text{3D}}, \Sigma_{\text{3D}}) = o \exp\left(-\frac{1}{2}(\mathbf{x} - \mu_{\text{3D}})^\top \mathbf{\Sigma_\text{3D}}^{-1}(\mathbf{x} - \mu_{\text{3D}})\right)
\end{equation}
Unlike implicit volumetric representations (e.g., NeRF \cite{mildenhall2021nerf}), 3DGS enables real-time rendering via explicit splat rasterization while preserving differentiability.

\subsubsection{Differentiable Rendering}

Following the formulation in \citet{kerbl20233d}, 3DGS rendering first projects 3D Gaussians to screen space. Given a Gaussian splat with 3D mean $\mu_\text{3D}$ and covariance $\mathbf{\Sigma}_\text{3D}$, its 2D projection is computed using the view matrix $\mathbf{W}$ and the projection Jacobian $\mathbf{J}$:
\[
\mu_\text{2D} = \text{proj}(\mathbf{W}\mu_\text{3D}), \quad \mathbf{\Sigma}_\text{2D} = \mathbf{J}\mathbf{W}\mathbf{\Sigma}_\text{3D}\mathbf{W}^\top\mathbf{J}^\top
\]
yielding the 2D Gaussian density function $\alpha_\text{2D}(\mathbf{u}|o,\mu_{\text{2D}}, \Sigma_{\text{2D}})$, where $\mathbf{u}$ denotes 2D screen-space coordinates and all parameters are projected to screen space.

Depth-sorted splats (indexed $1, \dots, N$) are alpha-composited to compute the final pixel color $C$. Let $c_i$ denote the view-dependent color of splat $i$, and $\alpha_i = \alpha_\text{2D}(\mathbf{u})$ (ignore $o, \mu_{\text{2D}}, \Sigma_{\text{2D}}$ for simplicity) as the transparency at pixel $\mathbf{u}$. The blending equations are:  
\begin{align}
T_i = \prod_{k=1}^{i-1} (1 - \alpha_k), \quad C = \sum_{i=1}^N T_i \alpha_i c_i \label{eq:C}
\end{align}
 
Where $T_i$ represents accumulated transmittance. To enable gradient-based optimization, the renderer computes derivatives of the loss $\mathcal{L}(C)$ with respect to splat parameters. The key gradients $\partial\mathcal{L}/\partial\alpha_i$ and $\partial\mathcal{L}/\partial c_i$ propagate to spatial parameters ($\mu_\text{3D}$, $\mathbf{\Sigma}_\text{3D}$) via chain rule differentiation. From \autoref{eq:C}, we can derive:  
\begin{align}
\frac{\partial\mathcal{L}}{\partial c_i} &= \frac{\partial\mathcal{L}}{\partial C} \alpha_i T_i, \label{eq:dL_dColor} \\
\frac{\partial\mathcal{L}}{\partial \alpha_i} &= \frac{\partial\mathcal{L}}{\partial C} \left(c_i T_i - \frac{\sum_{k=i+1}^N T_k \alpha_k c_k}{1 - \alpha_i}\right)  \label{eq:dL_dAlpha}
\end{align}  

The gradient computation supports both front-to-back($i=1,\dots,N$) and back-to-front ($i=N,\dots,1$) processing of depth-sorted splats. 

Although Gaussian splats have infinite theoretical support (since $\alpha_\text{2D}(\mathbf{u}) > 0$ for all $\mathbf{u}$), 
practical implementations discard marginal contributions with $\alpha_\text{2D}(\mathbf{u}) < 1/255$, 
a threshold aligned with 8-bit color precision. This early culling applies to both color blending and gradient computation. 

\subsubsection{Tile-based Rasterization}

The differentiable tile-based rasterizer proposed by \citet{kerbl20233d} partitions the screen into $16 \times 16$ tiles. For each splat, intersecting tiles are computed according to the splat boundary, and buffers are dynamically allocated to store splat-tile pairs. These pairs are sorted by tile ID and depth to generate per-tile depth-ordered splat lists. During the forward pass, pixel colors are computed in front-to-back order as \autoref{eq:C}.

In the backward pass, splats are processed in reversed depth order. 
Initial gradient computation focuses on per-pixel $\partial\mathcal{L}/\partial c_i$ and $\partial\mathcal{L}/\partial\alpha_i$, followed by propagation to intermediate derivatives, such as $\partial\mathcal{L}/\partial\mathbf{\Sigma_{\text{2D},i}}^{-1}$, $\partial\mathcal{L}/\partial o_i$ and $\partial\mathcal{L}/\partial \mu_{\text{2D},i}$. These gradients are then aggregated per-splat via atomic operations. High-dimensional parameter gradients (3D positions $\mu_\text{3D}$, SH coefficients, etc.) are processed in a dedicated subsequent pass to maintain performance.

\subsubsection{T-Culling}

\citet{kerbl20233d} also implements an optimization termed T-Culling, which early terminates front-to-back splat traversal in the forward pass when the accumulated transparency product $T_i$ drops below $0.0001$, with its results stored and reused to accelerate the backward pass as well. Leveraging the monotonic decay of $T_i$ with depth order, this heuristic skips deeper splats whose contributions to pixel color and gradients become negligible. This optimization has been widely adopted in subsequent tile-based rasterizers as in gsplat\cite{ye2024gsplatopensourcelibrarygaussian}.
Though being straightforward for tile-based methods, T-Culling is incompatible with hardware 3DGS rasterization when employing graphics pipeline fixed-function blending, as GPU's fixed-function blending intrinsically lacks support for such optimization.

\subsection{Improving 3DGS Rendering Performance}

Recent advances in accelerating 3DGS rendering span multiple directions, including compressing and pruning 3DGS models \cite{fan2024lightgaussian,girish2024eagles,fang2024mini,lee2024compact,niemeyer2024radsplat} and designing specialized hardware architectures for 3DGS rendering \cite{feng2024potamoi,lin2025metasapiens}. 

Parallel efforts focus on optimizing tile-based rasterizers. Works like GSCore \cite{lee2024gscore}  and FlashGS \cite{feng2024flashgs} enhance splat-tile intersection computation through precise geometric algorithms, significantly reducing noneffective splat-tile pairs compared to the naive AABB-tile approach. This optimization improves both forward/backward pass performance while reducing memory consumption. GSCore further adopts coarse-grained sorting and subtile skipping to optimize mobile deployment. Balanced3DGS\cite{gui2024balanced} on the other hand resolves GPU workload imbalance in tile-based methods caused by non-uniform splat distributions through dynamic tile-to-thread-group task allocation and intra-group splat-wise parallelism. DISTWAR \cite{durvasula2023distwar,durvasula2025arc} targets backward pass efficiency by minimizing atomic operation stalls. Its warp-level gradient pre-accumulation strategy aggregates per-thread gradients within a warp before committing to global memory via \texttt{atomicAdd}, achieving up to $5.7\times$ speedups in gradient computation. 

Despite their performance gains, these tile-based methods still suffer from inherit scalability issues: the need for dynamic memory management and a worst-case $\mathcal{O}(N \times M)$ memory requirement ($N$ splats and $M$ tiles) that becomes prohibitive at scale.

In constrast, 3DGS hardware rasterizers operates under fixed $\mathcal{O}(N)$ memory budgets. These approaches project splats as 2D quads through eigendecomposition of $\mathbf{\Sigma}_\text{2D}$, leveraging hardware interpolation and fixed-function blending to efficiently compute $\alpha$ and $\alpha$-composite colors respectively. Notable implementations include the WebGL-based renderer \cite{kevin2023splat}, OpenGL-powered frameworks \cite{fast_gauss} (reporting $5-10\times$ speedups over unoptimized tile-based rasterizers despite hardware blending limitations precluding T-Culling), and the Unity engine integration \cite{aras_unitygs}. These implementations lack frustum culling, a standard optimization in production-grade renderers, leaving measurable performance gains untapped. More critically, none implement backward passes, preventing their application to differentiable rendering tasks.

The core challenge for differentiable hardware rasterization lies in fragment shaders' inability for depth-ordered splat traversal needed by per-pixel gradient computation. 
Existing attempts to address this challenge rely on approximations that compromise mathematical fidelity relative to exact tile-based gradient computation. The depth peeling solution proposed by \citet{xu20244k4d} renders sequential depth layers to approximate full splat traversal, at the cost of omitting gradients from occluded splats beyond the layer limit ($\leq15$ in practice). \citet{kheradmand2025stochasticsplats} circumvents ordered traversal through Monte Carlo gradient estimation based on stochastic transparency \cite{enderton2010stochastic}, achieving hardware-compatible implementation at the cost of noisy gradients.

\section{Method}

\subsection{Gradient Computation via Programmable Blending}

In this section, we address the fundamental barrier to differentiable rasterization: fragment shaders' inability to access overlapping splat sequences. Our approach exploits programmable blending to enable correct per-pixel gradient computation.

\subsubsection{Programmable Blending}

Programmable Blending is a hardware rasterization technique that overcomes the predefined constraints of traditional fixed-function blending, enabling custom color mixing logic to be implemented in shader code. This capability, supported through graphics API extensions on modern GPUs, provides two essential features: (1) race-condition-free operations on per-pixel resources in fragment shaders, and (2) execution order guarantees which match the rasterization sequence for resource access.

Tile-based architecture GPUs prevalent in mobile platforms natively support programmable blending\cite{armbestpractice_mr,metalfeature}. These GPUs guarantee that fragment shaders within a tile execute in strict rasterization order, requiring only framebuffer synchronization to make programmable blending available. For example, within the Vulkan graphics API \cite{bailey2019introduction}, this capability is formally specified through the \texttt{VK\-\_EXT\-\_shader\-\_tile\-\_image} and \texttt{VK\-\_EXT\-\_rasterization\-\_order\-\_attachment\-\_access} extensions, which permit safe read-modify-write operations on the same attachment in fragment shaders while eliminating explicit memory barrier requirements.

Desktop GPUs programs can implement programmable blending via the \texttt{VK\_EXT\_fragment\_shader\_interlock} Vulkan extension, which defines a critical section through \texttt{begin\-Invocation\-Interlock()} and \texttt{end\-Invocation\-Interlock()} calls, enabling safe access to pixel-local data structures in fragment shaders. Execution order can be further configured within these critical sections as rasterization-ordered. However, unlike tile-based mobile GPUs that guarantee rasterization-order execution, most desktop GPUs employ out-of-order fragment shader scheduling for improved hardware utilization. This architectural approach requires fragment shader interlocks to enforce synchronization via pipeline stalls or execution order constraints, incurring measurable performance overhead.

\subsubsection{Front-to-Back Gradient Computation}

Programmable blending enables persistent per-pixel state management, making per-pixel gradient computation within fragment shaders feasible. While both front-to-back ($1 \to N$) and back-to-front ($N \to 1$) traversal orders are valid for gradient computation, we empirically find that front-to-back traversal better aligns with T-Culling optimizations. This stems from hardware blending's limitations in forward passes (not able to apply T-Culling), thereby constraining the optimization exclusively to backward passes. A back-to-front traversal in backward pass would require dividing out culled $1-\alpha_i$ terms for $T_i$ values needed by T-Culling, leading to extra memory accesses. Front-to-back traversal avoids this by directly computing $T_i$ through incremental updates.  

The front-to-back gradient computation proceeds as follows. Firstly, both \autoref{eq:dL_dColor} and \autoref{eq:dL_dAlpha} depend on the following recursively updated transmittance:  
\begin{align}
T_{i+1} = T_i (1 - \alpha_i), \quad T_1 = 1 \label{eq:T_update}
\end{align}
Additionally, the summation term in \autoref{eq:dL_dAlpha} requires recursive computation. We define an auxiliary sequence $C'_i$:  
\begin{align}
C'_i &= \sum_{k=i}^N T_k\alpha_k c_k, \quad C'_1 = C, \quad C'_{i+1} = C'_i - T_i\alpha_i c_i \label{eq:C'_update} 
\end{align}
Substituting \autoref{eq:C'_update} into \autoref{eq:dL_dAlpha} yields the reformed gradient expression:

\begin{align}
\frac{\partial\mathcal{L}}{\partial\alpha_i} = \frac{\partial\mathcal{L}}{\partial C} \cdot \frac{c_i T_i - C'_i}{1 - \alpha_i} \label{eq:dL_dAlpha_C'}
\end{align}

Using these relations, we compute $\partial\mathcal{L}/\partial c_i$ in \autoref{eq:dL_dColor} and $\partial\mathcal{L}/\partial \alpha_i$ in \autoref{eq:dL_dAlpha_C'} through front-to-back recurrence of $T_i$ and $C'_i$, serving as the foundation for deriving any other gradients.  

\subsubsection{Algorithm}  

Our backward pass draws splats in front-to-back order, and \autoref{alg:propagateGradient} illustrates our gradient computation approach within the fragment shader, enabled by programmable blending through \texttt{VK\_EXT\_fragment\_shader\_interlock} (targeting desktop GPUs). During the rasterization-ordered critical section enforced by the interlocks, we maintain and retrieve values of $C'_i$ and $T_i$ by reading/writing a dedicated texture $\text{tex}_{C',T}$. The shader then loads $\partial\mathcal{L}/\partial C$ from a read-only texture $\text{tex}_{\partial\mathcal{L}/\partial C}$, then computes $\partial\mathcal{L}/\partial c_i$ and $\partial\mathcal{L}/\partial\alpha_i$ using analytical derivatives derived from \autoref{eq:dL_dColor} and \ref{eq:dL_dAlpha_C'}. Similar to \citet{kerbl20233d}, these gradients are subsequently propagated to intermediate splat parameters $\theta_i$ (e.g., 2D positions, colors) via the chain rule, denoted as $\partial\mathcal{L}/\partial \theta_i$.  

Additionally, we maintain a boolean \texttt{cullingFlag} to track fragment rejection status, which combines two culling criteria: splat boundary culling ($\alpha_i < 1/255$) and T-Culling ($T_i < 0.0001$). Fragments marked by \texttt{cullingFlag} skip all gradient computations, reducing both texture accesses and arithmetic operations.

\newcommand{\highlight}[1]{\textcolor{blue!80!black}{\texttt{#1}}}

\begin{algorithm}[htbp]
\caption{\textsc{ComputeGradient}}\label{alg:propagateGradient}
\begin{algorithmic}[1]
\REQUIRE pixel coordinate $\mathbf{u}$, fragment alpha $\alpha_i$, splat color $c_i$, texture $\text{tex}_{C',T}$ initialized as $(C'_1, T_1) = (C, 1)$, readonly texture $\text{tex}_{\partial\mathcal{L}/\partial C}$ storing $\partial\mathcal{L}/\partial C$
\ENSURE splat parameter gradients $\partial\mathcal{L}/\partial\theta_i$, cullingFlag
\STATE $\text{cullingFlag} \gets \alpha_i < 1 / 255$ \COMMENT{Cull against splat boundary}
\item[]
\STATE \highlight{beginInvocationInterlock()} \COMMENT{Critical section}
\IF{\textbf{not} cullingFlag}
  \STATE $(C'_i, T_i) \gets \texttt{textureLoad}(\text{tex}_{C',T}[\mathbf{u}])$
  \IF[T-Culling]{$T_i < 0.0001$}
    \STATE $\text{cullingFlag} \gets \textbf{true}$
  \ELSE
    \STATE $(C'_{i+1}, T_{i+1}) \gets (C'_i - T_i \alpha_i c_i, T_i (1 - \alpha_i))$ \COMMENT{\autoref{eq:C'_update}, \ref{eq:T_update}}
    \STATE $\texttt{textureStore}(\text{tex}_{C',T}[\mathbf{u}], (C'_{i+1}, T_{i+1}))$
  \ENDIF
\ENDIF
\STATE \highlight{endInvocationInterlock()}
\item[]  \label{alg:line:earlyTerminate}
\IF{cullingFlag}
  \STATE $\partial\mathcal{L}/\partial \theta_i \gets 0$
\ELSE
  \STATE $\partial\mathcal{L}/\partial C \gets \texttt{textureLoad}(\text{tex}_{\partial\mathcal{L}/\partial C}[\mathbf{u}])$
  \STATE $\partial\mathcal{L}/\partial c_i \gets \partial\mathcal{L}/\partial C \cdot \alpha_i T_i$ \COMMENT{\autoref{eq:dL_dColor}}
  \STATE $\partial\mathcal{L}/\partial \alpha_i \gets \partial\mathcal{L}/\partial C \cdot (c_i T_i - C'_i) / (1 - \alpha_i)$ \COMMENT{\autoref{eq:dL_dAlpha_C'}}
  \STATE $\partial\mathcal{L}/\partial \theta_i \gets \text{bwd\_diff}(\partial\mathcal{L}/\partial c_i, \partial\mathcal{L}/\partial \alpha_i)$
\ENDIF

\end{algorithmic}
\end{algorithm}

\subsection{Hybrid Gradient Reduction}

With per-pixel splat parameter gradients computed, a complete backpropagation can then be achieved through splat-wise accumulation. As direct use of \texttt{atomicAdd} operations incurs substantial performance overhead due to global memory contention, we introduce a hybrid gradient reduction strategy leveraging fragment shader parallelism through quad-local and subgroup-level reductions to address this limitation.

\subsubsection{Fragment Shader Subgroups and Quads}

Prior to detailing our method, we briefly clarify the architectural foundations of subgroups and quads in modern GPUs.  Modern GPU architectures employ two key execution models for fragment shaders:

\textbf{Subgroups} (warps/wavefronts): Minimal schedulable thread blocks (32/64 threads on NVIDIA/AMD GPUs respectively) supporting intra-group data exchange via operations like \texttt{subgroupShuffle}.

\textbf{Quads}: $2\times2$ fragment blocks enabling derivative calculations, with helper invocations (detectable via \texttt{gl\_HelperInvocation}) handling primitive boundary conditions.

Fragment shaders rely exclusively on these units for thread communication due to lack of shared memory, constraining gradient aggregation to intra-quad and intra-subgroup operations.

\subsubsection{Quad Reduction}

To mitigate performance bottlenecks from frequent \texttt{atomic\-Add} operations, we propose aggregating gradients computed across multiple fragment shader threads using quad and subgroup intrinsics, followed by sparse \texttt{atomicAdd} calls from selected threads to update global memory buffers.  

A straightforward approach involves quad reduction: Since hardware rasterization processes splat primitives at quad granularity, all threads within a quad inherently belong to the same splat. By leveraging quad intrinsics (e.g., Vulkan's \texttt{subgroupQuadSwapVertical}) to accumulate gradients within the quad, we perform partial sums and delegate a single thread per quad to execute \texttt{atomicAdd}. This strategy reduces \texttt{atomicAdd} calls to below $30\%$ of the baseline approach and achieves 3.05$\times$ speedup (see \autoref{tab:reductionPerf}).

\subsubsection{Splat-Subgroup Cohesion and Subgroup Reduction} \label{sec:subgroupReduction}

To further reduce \texttt{atomicAdd} invocations, we explore subgroup gradient reduction. While there exists no guarantee that fragments within a subgroup originate from the same splat, we propose the Splat-Subgroup Cohesion Hypothesis: GPUs exhibit a statistical tendency to schedule fragments from the same splat (rendered as two adjacent triangle primitives) within a single subgroup, leveraging spatial locality for memory access efficiency during rasterization. Although not universally guaranteed, this hypothesis facilitates effective subgroup-level gradient reduction in most practical scenarios.

\begin{figure}[htbp]
  \centering
  {
    \newcommand{\cfbox}[2]{%
      \colorlet{currentcolor}{.}%
      {\color{#1}%
      \fbox{\color{currentcolor}#2}}%
    }
    \setlength{\fboxsep}{-2px} 
    \setlength{\fboxrule}{2px}
    \setlength{\tabcolsep}{2pt}
    \begin{minipage}{\dimexpr\linewidth/3\relax}
      \begin{picture}(\linewidth,\linewidth)
        \setlength{\unitlength}{\dimexpr\linewidth/720\relax}
        \put(0,0){\includegraphics[width=\linewidth]{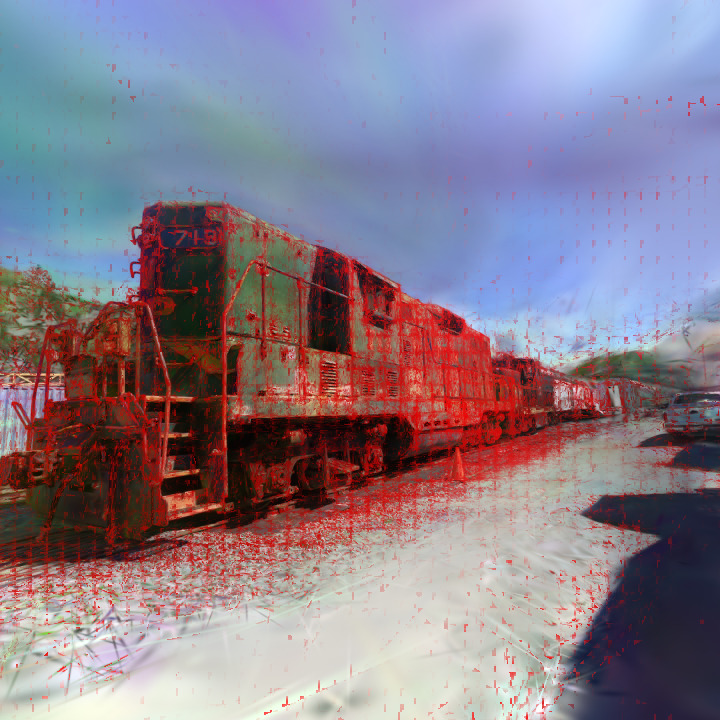}}
        \put(70,400){\linethickness{2px}\color{orange}\framebox(80,80){}}%
        \put(430,210){\linethickness{2px}\color{cyan}\framebox(80,80){}}%
      \end{picture}
    \end{minipage}\hfill
    \begin{minipage}{\dimexpr\linewidth/3\relax}
      \cfbox{orange}{\includegraphics[width=\linewidth]{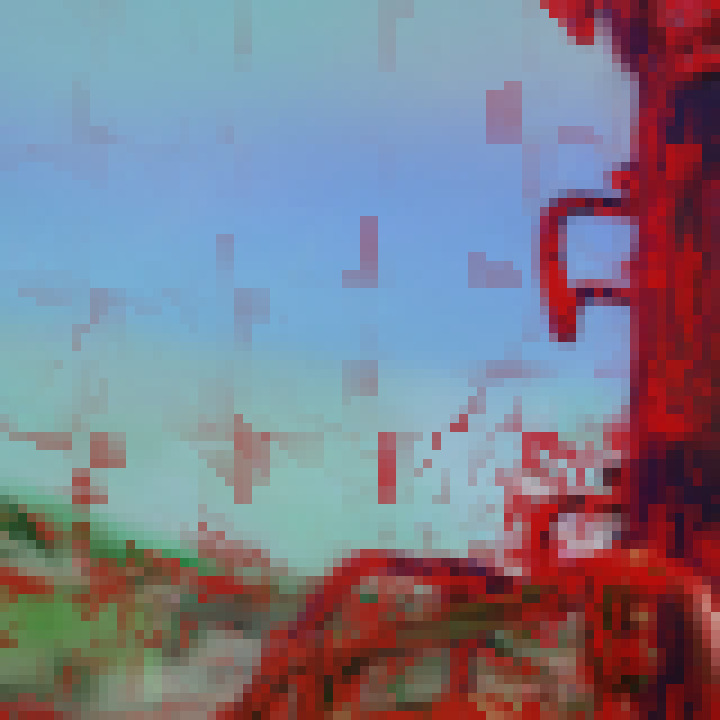}}
    \end{minipage}\hfill
    \begin{minipage}{\dimexpr\linewidth/3\relax}
      \cfbox{cyan}{\includegraphics[width=\linewidth]{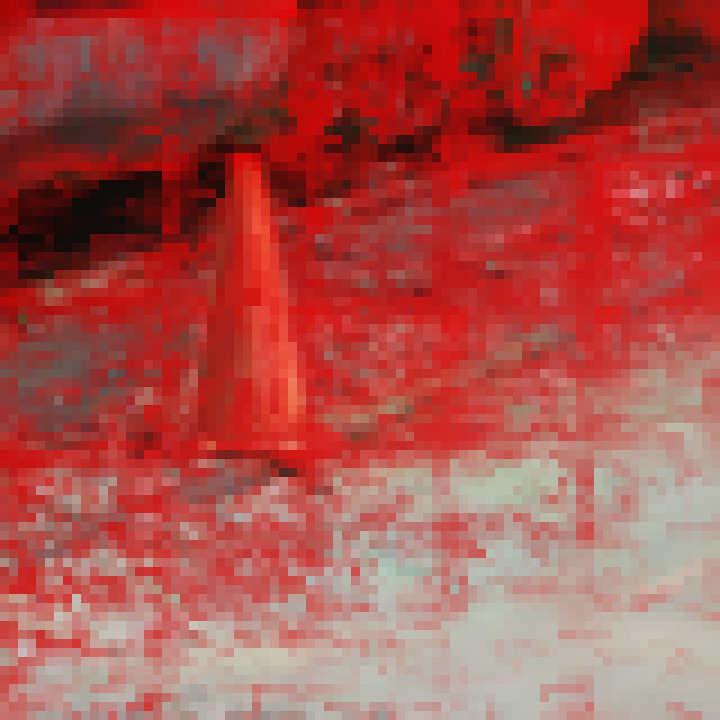}}
    \end{minipage}
  }

  \caption{Qualitative validation of splat-subgroup coherency in the \textsc{train} scene from the Tanks and Temples dataset \cite{Knapitsch2017}, rendered at $720\times720$ resolution. Splat-incoherent fragments (rendered in red) account for a small proportion and cluster on dense, small splats.}
  \label{fig:cohesion}
\end{figure}

We validate this hypothesis through preliminary experiments on an RTX 4080 GPU. Qualitative analysis of splat-subgroup coherency (\autoref{fig:cohesion}) shows that splat-incoherent fragments (processed in subgroups containing fragments from multiple splats) mainly occur with small splats on screen, while larger splats predominantly produce splat-coherent fragments. This indicates high splat-coherency under typical rendering conditions. Quantitative measurements (\autoref{fig:cohesionRate}) further confirm these findings.

These results confirm that our Splat-Subgroup Cohesion Hypothesis holds robustly on RTX 4080 GPUs. While architecture-specific, the observed pixel locality patterns suggest potential applicability to other modern GPUs, requiring further cross-platform validation.

Building on this hypothesis, our subgroup reduction pipeline first assesses splat cohesion using Vulkan's \texttt{subgroupAllEqual} intrinsic to verify uniform splat IDs across active threads. For cohesive subgroups, gradients are aggregated via \texttt{subgroupAdd}, with a single \texttt{atomicAdd} operation performed by a selected active lane. 

Standalone subgroup reduction reduces \texttt{atomicAdd} invocations to 6\% of baseline levels, and combining it with quad reduction further decreases the ratio to approximately 5\%.

\subsubsection{Balancing Threshold in Subgroup Reduction}

While minimizing \texttt{atomicAdd} invocations improves performance, excessive aggregation via subgroup reduction risks overloading streaming multiprocessors (SMs) with computation, creating an imbalance between SM workloads and L2 atomic unit (ROP) utilization that incurs performance penalties\cite{durvasula2023distwar,durvasula2025arc}. Inspired by \citet{durvasula2023distwar,durvasula2025arc}, we introduce a balancing threshold $X$ and restrict subgroup reduction to cases where the number of active threads requiring gradient aggregation exceeds the threshold. The optimal balancing threshold $X$ is hardware-and-scene-dependent and need to be selected through performance profiling.

\subsubsection{Algorithm} 

Our hybrid gradient reduction method, combining quad-based and subgroup-based reduction strategies, is detailed in \autoref{alg:reduceGradient}. This approach employs Vulkan GLSL intrinsics for subgroup and quad operations. The algorithm initiates by filtering out noneffective threads through a \texttt{reduceFlag}, which combines \texttt{cullingFlag} (from \autoref{alg:propagateGradient}) and hardware-generated \texttt{gl\_HelperInvocation}. The \texttt{reduceFlag} excludes helper invocations (which cannot perform memory writes or framebuffer updates) and fragments previously discarded by splat boundaries or T-Culling. A bitmask \texttt{reduceMask}, constructed via \texttt{subgroupBallot}, identifies active threads eligible for gradient aggregation.  

Then, the method dynamically selects between subgroup and quad operations based on runtime conditions. The algorithm first evaluates subgroup reduction feasibility by checking splat ID coherence across active threads and verifying that the valid fragment count, measured via \texttt{subgroupBallotBitCount}, exceeds the balancing threshold $X$. When these criteria are met, gradients are aggregated within the subgroup. If unsatisfied, the method falls back to quad reduction, aggregating gradients across $2\times2$ pixel blocks, and then restricting atomic updates to quad-local threads by zeroing out \texttt{reduceMask} bits for threads outside the current quad. The quad reduction fallback acts as a safety net for cases where subgroup coherence cannot be guaranteed, preserving performance across diverse scene complexities. 

Finally, the least-significant thread in the current quad or subgroup reduction scope (identified via \texttt{subgroup\-Ballot\-Find\-LSB}) performs a single \texttt{atomicAdd}, accumulating the reduced gradients to global memory. 

This dual-strategy approach is derived through rigorous profiling across varying 3DGS workloads, with empirical validation of its efficacy provided in \autoref{tab:reductionPerf}.

\begin{algorithm}[htbp]
\caption{\textsc{ReduceGradient}}\label{alg:reduceGradient}
{
\renewcommand{\algorithmicdo}{}
\renewcommand{\algorithmicwhile}{\textbf{function}}
\renewcommand{\algorithmicendwhile}{\algorithmicend\ \algorithmicwhile}
\begin{algorithmic}[1]
\REQUIRE splat ID $i$, splat parameter gradients $\partial\mathcal{L}/\partial\theta_i$, cullingFlag
\ENSURE gradients accumulated in buffer $\text{buf}_{\partial\mathcal{L}/\partial\theta}$, initialized as $0$
\STATE $\text{reduceFlag} \gets \textbf{not } \space (\text{cullingFlag} \textbf{ or } \texttt{gl\_HelperInvocation})$

\STATE $\text{reduceMask} \gets \texttt{subgroupBallot}(\text{reduceFlag})$
\IF[Subgroup reduction]{$\texttt{subgroupAllEqual}(i) \textbf{ and } $\\$ \quad\texttt{subgroupBallotBitCount}(\text{reduceMask}) \geq \text{X}$\\} 
  \STATE $\partial\mathcal{L}/\partial\theta_i \gets \texttt{subgroupAdd}(\partial\mathcal{L}/\partial\theta_i)$
\ELSE[Quad reduction]
  \STATE $\partial\mathcal{L}/\partial\theta_i \gets \textsc{quadAdd}(\partial\mathcal{L}/\partial\theta_i)$
  \STATE $\text{reduceMask} \gets \text{reduceMask} \mathbin{\&} \texttt{gl\_SubgroupEqMask}$
  \STATE $\text{reduceMask} \gets \textsc{quadOr}(\text{reduceMask})$
\ENDIF

\IF{$\texttt{gl\_SubgroupInvocationID} \textbf{ equals }$\\$\quad\texttt{subgroupBallotFindLSB}(\text{reduceMask})$}
  \STATE $\texttt{atomicAdd}(\text{buf}_{\partial\mathcal{L}/\partial\theta}[i], \partial\mathcal{L}/\partial\theta_i)$
\ENDIF
{\color{gray}
\WHILE{\textsc{quadAdd}($a$)}
  \STATE $a \gets a + \texttt{subgroupQuadSwapHorizontal}(a)$
  \STATE $a \gets a + \texttt{subgroupQuadSwapVertical}(a)$
  \RETURN $a$
\ENDWHILE
\WHILE{\textsc{quadOr}($b$)}
  \STATE $b \gets b \mathbin{|} \texttt{subgroupQuadSwapHorizontal}(b)$
  \STATE $b \gets b \mathbin{|} \texttt{subgroupQuadSwapVertical}(b)$
  \RETURN $b$
\ENDWHILE}
\end{algorithmic}
}
\end{algorithm}

\subsection{Mixed Precision Rendering}

Besides backward pass optimizations, forward pass performance also determines our rasterizer's overall efficiency. Our experiments identify render target texture formats as the primary performance determinant in 3DGS forward pass execution. Notably, fixed-function blending with float32 proves particularly inefficient, even underperforming fragment interlock approaches. In contrast, reduced-precision formats (float16, unorm16, unorm8) demonstrate the superior performance of GPU rasterization and blending, with unorm8 delivering the fastest rendering speeds, while float16 and unorm16 achieve comparable runtime efficiency (see \autoref{tab:perf}).

To balance performance and numerical consistency, we adopt a mixed precision rendering strategy that unifies precision across both forward and backward passes: the render target (storing colors and $T$'s) and the per-pixel state texture $\text{tex}_{C',T}$ (in \autoref{alg:propagateGradient}) adopt reduced-precision formats, while intermediate calculations preserve full float32 precision. This ensures numerically accurate gradient computation while eliminating precision-mismatch artifacts during backpropagation. 
In addition, when using unorm formats, the color value $c_i$ must be saturated within $[0,1]$ to prevent overflow.

Systematic evaluation of gradient accuracy (\autoref{tab:precision}) demonstrates that 16-bit formats (float16 and unorm16) introduce relatively minor errors, while unorm8 exhibits significantly larger deviations. Thus, we suggest prioritizing 16-bit formats as they deliver substantial speedup over full-precision formats while maintaining accuracy. Although unorm8 offers the fastest rendering speed, the substantially higher error levels may adversely affect 3DGS training stability.


\section{Experiments}

We evaluated our method on the MipNeRF360 dataset \cite{barron2022mip} using pretrained 3D Gaussian Splatting models (30k iterations) from the original implementation \cite{kerbl20233d}. The rendering of 3DGS models defaults to the dataset's native image resolutions. Numerical precision (across float32 and lower-precision formats) and backward pass performance are measured using stochastic image gradients uniformly distributed in $[-1,1]$. 

Our comparative benchmarks include: 

\textbf{3DGS baseline}: The original tile-based 3DGS CUDA rasterizer \cite{kerbl20233d}.

\textbf{DISTWAR}: Tile-based variant with optimized backward pass for gradient accumulations, while preserving forward pass performance and memory footprint \cite{durvasula2023distwar,durvasula2025arc}.

\textbf{3DGS float16}: Modified baseline 3DGS rasterizer using float16 image formats and pixel operations, simulating our mixed-precision approach's storage/computational loads.

All experiments run on an NVIDIA RTX 4080 Laptop GPU, with Vulkan and CUDA performance metrics rigorously measured using low-level timers (\texttt{vkQueryPool} for Vulkan, \texttt{cudaEvent} for CUDA) to ensure accuracy. 

\subsection{Implementation Details}

Our code implementation utilizes C++, GLSL, and Vulkan. We develop a custom version of the \textit{Onesweep} sorting algorithm \cite{adinets2022onesweep,merrill2016single} in the forward pass, enabling dynamic-length splat sorting via non-blocking indirect dispatch to support GPU-driven frustum culling, maximizing throughput. An optimization for backward pass occurs after \autoref{alg:line:earlyTerminate} in \autoref{alg:propagateGradient}, where we implement early termination (via \textbf{return} statement) when all fragments in the current quad are rejected (determined by \texttt{subgroup\-Quad\-All}(cullingFlag)), which reduces SM pressure while preserving compatibility with hybrid reduction's quad-level fallback.

\subsection{Splat-Subgroup Cohesion Rate}

Before evaluating performance metrics, we first present quantitative validation of the Splat-Subgroup Cohesion Hypothesis discussed in \autoref{sec:subgroupReduction}. We measure cohesion rates (incoherent fragments are depicted in red) across varying scenes (from MipNeRF360 dataset\cite{barron2022mip}), camera viewpoints, and resolutions, with results visualized in \autoref{fig:cohesionRate}.

We observe that scenes with higher geometric complexity (e.g., \textsc{bicycle}, \textsc{garden}, \textsc{stump}) exhibit lower cohesion rates than simpler scenes (\textsc{room}, \textsc{counter}, \textsc{kitchen}) at equivalent resolutions, while cohesion rates scale monotonically with rendering resolution, rising from about $70\%$ at $1024\times512$ to approximately $90\%$ at $2048\times1024$, ultimately approaching $100\%$ at $4096\times2048$. 

As geometric complexity decreases and rendering resolution increases, individual splats cover more pixels in screen space, leading to the above correlation. The expanded splat coverage increases the likelihood of processing multiple pixels of a splat primitive within the same fragment shader subgroup, thereby statistically elevating cohesion rates. These measurable dependencies can thus validate our Splat-Subgroup Cohesion Hypothesis.

Notably, splat-incoherent fragments exhibit grid-aligned distributions, suggesting GPU-specific rasterization thread scheduling patterns.

\subsection{Gradient Reduction Performance}

We begin performance evaluation with gradient reduction methods, testing (1) naive \texttt{atomicAdd}, (2) quad reduction, (3) subgroup reduction, and (4) hybrid reduction on the MipNeRF360 dataset. Our analysis focuses on backward pass rasterization performance to validate hybrid reduction's efficacy (see \autoref{tab:reductionPerf}). For subgroup and hybrid methods, we assessed the balancing threshold's ($X$) impact while monitoring two factors: the proportion of fragments using \texttt{atomicAdd} (atomic add rate) and runtime performance metrics.  

\begin{table}[htbp]
  \caption{Backward pass rasterization performance and atomic-add rate (average) comparison of gradient reduction methods on MipNeRF360 dataset\cite{barron2022mip}.}
  \label{tab:reductionPerf}
  \centering
  {
    \setlength{\tabcolsep}{8pt}
    \begin{tabular*}{\linewidth}{l*{3}{c}}
      \toprule
      \multirow{2}{*}{Method} & \multicolumn{3}{c}{Rasterization} \\

      \cmidrule(lr){2-4}

      & Atomic-Add & Perf. & Speedup \\
      \midrule
      Naive & 1.000 & 462.34 ms & 1$\times$ \\
      Quad & 0.287 & 151.46 ms & 3.05$\times$ \\
      Subgroup (no $X$) & 0.060 & 45.70 ms & 10.12$\times$ \\
      Subgroup ($X = 4$) & 0.064 & 45.34 ms & 10.20$\times$ \\
      Hybrid (no $X$) & \textbf{0.051} & 44.97 ms & 10.28$\times$ \\
      Hybrid ($X = 8$) & 0.055 & \textbf{44.27 ms} & \textbf{10.44}$\times$ \\
      \bottomrule
    \end{tabular*}
  }
  \vspace{-1em}
\end{table}

Results demonstrate that hybrid reduction achieves $10\times$ speedup in backward pass rasterization compared to naive \texttt{atomic\-Add}, while outperforming standalone quad and subgroup reductions. Employing an optimal balancing threshold ($X = 4$ for subgroup reduction and $X = 8$ for hybrid reduction) further enhances performance despite slightly increasing the atomic-add rate.

\subsection{Overall Performance}

We conducted performance measurements on the MipNeRF360 dataset across various render target formats, comparing our method with tile-based methods (see \autoref{tab:perf}).

\begin{table*}[htbp]
  \centering
  \caption{Performance evaluation results on MipNeRF360 dataset \cite{barron2022mip}. Metrics include rasterization-only and overall throughput (average) for forward/backward passes as well as end-to-end (forward + backward) latency, in which the overall forward/backward duration comprising rasterization, preprocessing, and sorting stages (excluding tile-based methods' memory allocation). Ranked results are color-coded: \colorbox{FirstPlace}{1st}, \colorbox{SecondPlace}{2nd}, \colorbox{ThirdPlace}{3rd} per metric column.}
  \label{tab:perf}

  {\setlength{\tabcolsep}{6pt}
  \begin{tabular*}{\linewidth}{l*{10}{c}}
    \toprule
    \multirow{3}{*}{Method} 
    & \multicolumn{4}{c}{Forward}
    & \multicolumn{4}{c}{Backward} 
    & \multicolumn{2}{c}{\multirow{2}{*}{End-to-End}} \\ 
    \cmidrule(lr){2-5} \cmidrule(lr){6-9}
    & \multicolumn{2}{c}{Rasterization} 
    & \multicolumn{2}{c}{Overall} 
    & \multicolumn{2}{c}{Rasterization} 
    & \multicolumn{2}{c}{Overall} 
    & \\ 
    \cmidrule(lr){2-3} \cmidrule(lr){4-5}
    \cmidrule(lr){6-7} \cmidrule(lr){8-9}
    \cmidrule(lr){10-11}
    & Perf. & Speedup & Perf. & Speedup 
    & Perf. & Speedup & Perf. & Speedup 
    & Perf. & Speedup \\ 
    \midrule
    3DGS baseline 
    & \multirow{2}{*}{16.74 ms} & \multirow{2}{*}{1$\times$} 
    & \multirow{2}{*}{36.06 ms} & \multirow{2}{*}{1$\times$} 
    & 141.3 ms & 1$\times$ & 143.7 ms & 1$\times$ 
    & 179.7 ms & 1$\times$ \\
    DISTWAR 
    & & & & 
    & \cellcolor{FirstPlace}35.83 ms & \cellcolor{FirstPlace}3.94$\times$ & \cellcolor{FirstPlace}38.34 ms & \cellcolor{FirstPlace}3.75$\times$ 
    & 74.40 ms & 2.42$\times$ \\
    3DGS float16 
    & 17.08 ms & 0.98$\times$ & 36.19 ms & 1.00$\times$ 
    & 140.3 ms & 1.01$\times$ & 142.7 ms & 1.01$\times$ 
    & 178.9 ms & 1.00$\times$ \\
    \cmidrule(lr){1-11}
    Ours (float32) 
    & 62.03 ms & 0.27$\times$ & 63.86 ms & 0.56$\times$ 
    & 48.67 ms & 2.90$\times$ & 51.04 ms & 2.82$\times$ 
    & 114.9 ms & 1.56$\times$ \\
    
    Ours (float16) 
    & \cellcolor{ThirdPlace}10.07 ms & \cellcolor{ThirdPlace}1.66$\times$ & \cellcolor{ThirdPlace}11.91 ms & \cellcolor{ThirdPlace}3.03$\times$ 
    & \cellcolor{ThirdPlace}44.27 ms& \cellcolor{ThirdPlace}3.19$\times$ & \cellcolor{ThirdPlace}46.65 ms & \cellcolor{ThirdPlace}3.08$\times$ 
    & \cellcolor{SecondPlace}58.56 ms & \cellcolor{SecondPlace}3.07$\times$ \\
    
    Ours (unorm16) 
    & \cellcolor{SecondPlace}10.06 ms & \cellcolor{SecondPlace}1.66$\times$ & \cellcolor{SecondPlace}11.89 ms & \cellcolor{SecondPlace}3.03$\times$ 
    & 45.05 ms & 3.14$\times$ & 47.44 ms & 3.03$\times$ 
    & \cellcolor{ThirdPlace}59.32 ms & \cellcolor{ThirdPlace}3.03$\times$ \\
    Ours (unorm8) 
    & \cellcolor{FirstPlace}6.29 ms & \cellcolor{FirstPlace}2.66$\times$ 
    & \cellcolor{FirstPlace}8.13 ms & \cellcolor{FirstPlace}4.44$\times$ 
    & \cellcolor{SecondPlace}42.77 ms & \cellcolor{SecondPlace}3.30$\times$ & \cellcolor{SecondPlace}45.12 ms & \cellcolor{SecondPlace}3.18$\times$ 
    & \cellcolor{FirstPlace}53.25 ms & \cellcolor{FirstPlace}3.37$\times$ \\
    \bottomrule
  \end{tabular*}}
  \vspace{-1em}
\end{table*}

Our method demonstrates superior backward pass and end-to-end performance over baseline 3DGS, achieving over 2.9$\times$ speedup in backward rasterization. This validates the effectiveness of hybrid reduction, though limitations of fragment shaders constraint backward pass performance compared to DISTWAR.

When using low-precision render targets (float16, unorm16, unorm8), our approach attains 3$\times$ speedup in forward pass compared to tile-based methods, thereby achieving improved performance compared with DISTWAR despite backward pass performance disadvantages.

In contrast, tile-based methods show negligible benefits from the mixed precision method (speedup $<1.01\times$ for float16), as they operate on image pixels using ALU and registers, with no repeated global image read or write --- float16 textures thus only marginally reduce memory bandwidth, and half-precision ALU operations for pixel calculations provide limited advantages.

Surprisingly, float32 render targets exhibit severely degraded forward rasterization performance (0.27$\times$ of tile-based methods) with backward passes outperforming forward ones. This reveals GPU optimization deficiencies for rasterization and blending on full-precision formats, directly motivating our mixed-precision design.

\subsection{Memory-Efficiency}

In addition to runtime performance, we perform comprehensive investigation into the memory behavior. \autoref{tab:memory} compares the memory footprint between our method and tile-based approaches (3DGS baseline) on the MipNeRF360 dataset, where the metrics represent the minimum required GPU memory allocation for processing the entire dataset. By leveraging hardware rasterization, our method inherently bypasses splat-tile pair sorting, achieving a 37$\times$ reduction in sorting-stage memory consumption,
thus enabling over 4$\times$ total memory savings compared to tile-based methods without specialized optimizations. Adopting low-precision texture formats provides additional memory savings through compressed storage. 

\begin{table}[htbp]
  \centering
  \caption{Memory consumption on MipNeRF360 dataset\cite{barron2022mip}. Metrics includes two components: sorting-phase requirements (data buffers and GPU sorting algorithm auxiliary buffers) and total memory footprint (all memory allocations excluding 3DGS parameters, rendering results, and their corresponding input/output gradient buffers).}
  \label{tab:memory}
  {\setlength{\tabcolsep}{4.5pt}
  \begin{tabular*}{\linewidth}{l*{4}{c}}
    \toprule

    \multirow{2}{*}{Method} & \multicolumn{2}{c}{Sorting Mem.} & \multicolumn{2}{c}{Overall Mem.} \\
    
    \cmidrule(lr){2-3} \cmidrule(lr){4-5}
    
    & Allocated & Reduction & Allocated & Reduction \\
    
    \midrule
    
    3DGS baseline & 3.93 GB & 1$\times$ & 5.09 GB & 1$\times$ \\

    \cmidrule(lr){1-5}

    Ours (float32) & \multirow{2}{*}{105 MB} & \multirow{2}{*}{37.4$\times$} & 1.23 GB & 4.14 $\times$ \\

    Ours (unorm8) &  &  & 1.02 GB & 4.99 $\times$ \\
    
    \bottomrule
  \end{tabular*}}
  \vspace{-1em}

\end{table}

\begin{table}[htbp]
  \centering
  \caption{Gradient computation errors (average) of low-precision formats (float16, unorm16, unorm8) in mixed-precision rendering relative to full-precision (float32) in the MipNeRF360 dataset\cite{barron2022mip}, measuring mean, scale, rotation, opacity, and spherical harmonics gradients for RMSE and Mean Relative Error (MRE) across three magnitude ranges.}
  \label{tab:precision}
  {
  \newcolumntype{K}[1]{>{\centering\arraybackslash}p{#1}}
  \begin{tabular*}{\linewidth}{l*{4}{K{1.4cm}}}
    \toprule

    \multirow{2}{*}{Format} & \multirow{2}{*}{RMSE} & \multicolumn{3}{c}{Mean Relative Error} \\

    \cmidrule(lr){3-5}

     & & $[10, \infty)$ & $[0.1, 10)$ & $[1\mathrm{e}{-3}, 0.1)$ \\

    \midrule

    float16 & \textbf{0.197} & \textbf{0.022} & 0.104 & 1.594 \\
    unorm16 & 0.569 & 0.031 & \textbf{0.029} & \textbf{0.100} \\
    unorm8 & 2.358 & 0.273 & 1.318 & 23.83 \\
    
    \bottomrule
  \end{tabular*}
  }
  \vspace{-1em}
\end{table}

\subsection{Mixed Precision Errors}

Lastly we analyze the gradient computation errors of three low-precision texture formats (float16, unorm16, unorm8) relative to the float32 baseline under mixed-precision rendering. 

Results (\autoref{tab:precision}) indicate that 16-bit float/unorm formats exhibit superior precision retention, demonstrating lower root-mean-square errors (RMSE) and mean relative errors (MRE). Both formats achieve MRE below 0.04 for magnitudes $\geq 10$, with RMSE values below 0.6. Notably, unorm16 exhibits higher RMSE (0.569) than float16 (0.197), while its MRE remains lower for magnitudes $< 10$. This discrepancy suggests that while float16 maintains accuracy, it exhibits sensitivity to small-magnitude gradients, whereas unorm16 demonstrates poorer global precision but better consistency across scales.

In contrast, unorm8 shows significantly degraded accuracy, with RMSE exceeding float16 by over an order of magnitude and MRE peaking at 23.83 within $[0.001, 0.1)$. These findings confirm that float16 and unorm16 preserve gradient accuracy comparable to full-precision (float32) computations, whereas unorm8 introduces substantial errors that could negatively impact the optimization and training stability of 3DGS.


\section{Conclusion and Future Work}

Our differentiable 3DGS hardware rasterization method enables per-pixel gradient computation through programmable blending, accelerated by hybrid synchronization primitives for gradient accumulation. To address GPU inefficiencies with float32 render targets, we implement 16-bit formats that preserves numerical fidelity while delivering significant performance gains---achieving over 3.03$\times$ end-to-end speedup compared to the original tile-based 3DGS implementation. The hardware rasterization paradigm inherently reduces memory consumption by over 4.1$\times$ compared to tile-based approaches, enabling resource-constrained 3DGS training.

Future works include extending this architecture to Gaussian Splatting variants like 2DGS\cite{huang20242d}, and exploring mobile deployment for on-device training. While mobile GPUs natively support programmable blending, architectural differences in hybrid reduction implementations require thorough validation, which is a key technical challenge for our cross-platform adaptation.


\bibliographystyle{ACM-Reference-Format}
\bibliography{main}

\begin{figure*}[htbp]
  \centering
  {
    \newcommand\Image[2]{
      \begin{picture}(\linewidth,.5\linewidth)
        \setlength{\unitlength}{\dimexpr\linewidth/200\relax}
        \put(0,0){\includegraphics[width=\linewidth]{#1}}
        \put(5,8){\colorbox{blue!30}{{#2\%}}}
      \end{picture}
    }
    \def\arraystretch{0.4}
    \setlength{\tabcolsep}{0.8pt}
    \newcolumntype{K}[1]{>{\centering\arraybackslash}p{#1}}
    \begin{tabular*}{\linewidth}{*{3}{K{.33\linewidth}}}
      \Large{$1024\times 512$} & 
      \Large{$2048\times 1024$} & 
      \Large{$4096\times 2048$} \\
      \addlinespace[1ex]

      \Image{rtx4080verbose/bicycle_1024x512}{65.4} &
      \Image{rtx4080verbose/bicycle_2048x1024}{85.9} &
      \Image{rtx4080verbose/bicycle_4096x2048}{96.6} \\

      \Image{rtx4080verbose/garden_1024x512}{57.6} &
      \Image{rtx4080verbose/garden_2048x1024}{82.5} &
      \Image{rtx4080verbose/garden_4096x2048}{95.9} \\

      \Image{rtx4080verbose/stump_1024x512}{67.2} &
      \Image{rtx4080verbose/stump_2048x1024}{88.3} &
      \Image{rtx4080verbose/stump_4096x2048}{97.5} \\

      \Image{rtx4080verbose/room_1024x512}{80.5} &
      \Image{rtx4080verbose/room_2048x1024}{94.0} &
      \Image{rtx4080verbose/room_4096x2048}{98.6} \\

      \Image{rtx4080verbose/counter_1024x512}{87.2} &
      \Image{rtx4080verbose/counter_2048x1024}{96.6} &
      \Image{rtx4080verbose/counter_4096x2048}{99.4} \\

      \Image{rtx4080verbose/kitchen_1024x512}{76.6} &
      \Image{rtx4080verbose/kitchen_2048x1024}{93.3} &
      \Image{rtx4080verbose/kitchen_4096x2048}{98.9} \\
    \end{tabular*}
  }

  \caption{Quantitative and visual analysis of splat-subgroup cohesion across scenes and resolutions in the MipNeRF360 dataset \cite{barron2022mip} (top to bottom: \textsc{bicycle}, \textsc{garden}, \textsc{stump}, \textsc{room}, \textsc{counter}, \textsc{kitchen}). Red highlights denote splat-incoherent fragments---those processed in subgroups containing mixed fragments from multiple splats. The cohesion rate (numerically labeled in each image’s lower-left corner) correlates inversely with red coverage area: extensive red indicates lower cohesion, reflecting frequent inter-splat fragment interleaving within GPU subgroups. Spatial patterns of red regions further expose GPU scheduling biases that generate splat-incoherent subgroups under specific rendering conditions.}
  \label{fig:cohesionRate}
\end{figure*}
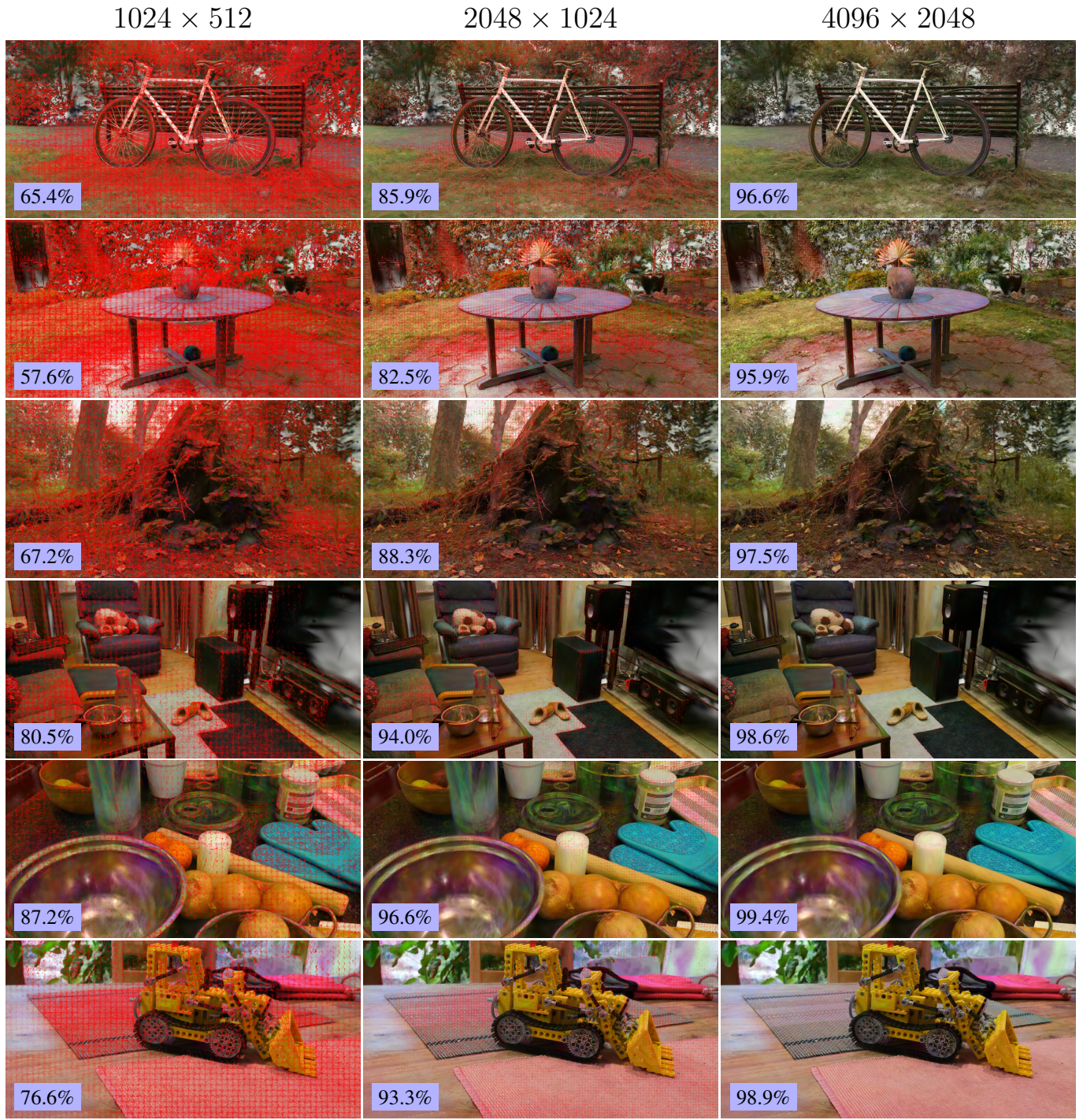



\end{document}